\documentclass[preprint,nofootinbib]{revtex4}

\usepackage{graphicx}

\begin{document}

\title{From de Sitter to de Sitter
\\{\small A non-singular inflationary universe driven by vacuum}}

\author{Saulo Carneiro\footnote{saulo@fis.ufba.br,
saulocarneiro@yahoo.com}}

\affiliation{Instituto de F\'{\i}sica, Universidade Federal da
Bahia, 40210-340, Salvador, BA, Brazil \\ International Centre for
Theoretical Physics, Trieste, Italy\footnote{Associate member}}

\begin{abstract}
A semi-classical analysis of vacuum energy in the expanding
spacetime suggests that the cosmological term decays with time, with
a concomitant matter production. For early times we find, in Planck
units, $\Lambda \approx H^4$, where $H$ is the Hubble parameter. The
corresponding cosmological solution has no initial singularity,
existing since an infinite past. During an infinitely long period we
have a quasi-de Sitter, inflationary universe, with $H \approx 1$.
However, at a given time, the expansion undertakes a phase
transition, with $H$ and $\Lambda$ decreasing to nearly zero in a
few Planck times, producing a huge amount of radiation. On the other
hand, the late-time scenario is similar to the standard model, with
the radiation phase followed by a dust era, which tends
asymptotically to a de Sitter universe, with vacuum dominating
again.
\end{abstract}

\maketitle

\begin{flushright}
{\small {\it ``Matter has to find a way to avoid the annihilation of
its volume"}\\ George Lema\^{\i}tre

\vspace{0.25cm}

{\it ``however, [...] `the beginning of the world' really
constitutes a beginning"}\\ Albert Einstein}
\end{flushright}

${ }$

The cosmological constant problem has been a theme of theoretical
discussion for decades, and has turned into a central point of
modern cosmology since recent observations suggested the existence
of a negative-pressure component in the cosmic energy content
\cite{Weinberg}.

The problem arises when we try to associate such a component with
the vacuum energy density predicted by quantum field theories. In
the case, for example, of a massless scalar field, the energy
density associated to its quantum fluctuations is given by
\begin{equation} \label{Lambdanua}
\Lambda_0 \approx \int_0^{\infty} \omega^3 d\omega.
\end{equation}

This divergent integral can be regularized by imposing a superior
cutoff $m$, leading to $\Lambda_0 \approx m^4$. This may also be
performed by introducing a bosonic distribution function in
(\ref{Lambdanua}),
\begin{equation}\label{Lambdareg}
\Lambda_0 \approx \int_0^{\infty}
\frac{\omega^3\,d\omega}{e^{\omega/m}-1}\approx m^4.
\end{equation}
The regularization procedure is thus equivalent to assume a thermal
distribution of vacuum fluctuation modes, at a characteristic
temperature $m$.

A natural choice for $m$ is the energy scale of QCD condensation,
the latest cosmological vacuum transition we know, since vacuum
fluctuations above this cutoff - which has the order of the pion
mass - would generate quark de-confinement. Unfortunately, even with
this value (many orders of magnitude below the Planck mass, also
usually taken as a natural cutoff), the obtained vacuum density is
around $40$ orders of magnitude higher than the presently observed
cosmological constant. That is the problem.

We should observe, however, that the above reasoning is based on
QFT in flat spacetime. In this case, the energy-momentum tensor
appearing in Einstein's equations must be zero, and, therefore,
(\ref{Lambdareg}) should be exactly canceled by a bare
cosmological constant. Such a cancelation should occur for any
vacuum contribution derived in flat spacetime.

Now, what would happen if we could calculate the vacuum density in
the expanding background? The regularized result would depend on the
curvature, and, after subtracting $\Lambda_0$, we should obtain a
renormalized, time-dependent cosmological term $\Lambda$, decaying
from high initial values to smaller ones, as the universe expands
\cite{Ozer,Schutzhold,GRF2005}. This renormalization is similar to
what happens in the Casimir effect, where the important thing is not
the vacuum density itself, but the difference between its values
inside a bounded region and in unbounded space.

The variation in the vacuum density leads, on the other hand, to
matter production, in order to preserve the conservation of total
energy implied by Einstein's equations. Indeed, in the realm of a
spatially homogeneous and isotropic spacetime, the Bianchi
identities lead to the conservation equation
\begin{equation} \label{continuidade}
\dot{\rho}_T + 3H(\rho_T+p_T)=0.
\end{equation}
Here, $H=\dot{a}/a$ is the Hubble parameter, while $\rho_T$ and
$p_T$ are, respectively, the total energy and pressure of cosmic
fluid. By introducing the matter density and pressure, and writing
$\rho_T = \rho_m + \Lambda$ and $p_T = p_m - \Lambda$, we
have\footnote{Since the vacuum has the same symmetry as spacetime,
its energy-momentum tensor has the form
$T_{\Lambda}^{\mu\nu}=\Lambda g^{\mu\nu}$, where $g^{\mu\nu}$ is the
metric tensor, and $\Lambda$ is a scalar function of coordinates (in
the FLRW spacetime, just a function of time). Therefore, it has the
same structure as for a perfect fluid in co-moving observers, with
$p_{\Lambda} = - \Lambda$.}
\begin{equation} \label{continuidade2}
\dot{\rho}_m+3H(\rho_m+p_m)=-\dot{\Lambda}.
\end{equation}
This shows that matter is not conserved - the decaying vacuum acting
as a source of entropy.

But how to evaluate the vacuum contribution in the expanding
spacetime? A possible answer is suggested by a semi-classical
analysis of the equation of motion of a minimally coupled massless
scalar field, $D^{\mu}D_{\mu}\phi=0$, where $D$ denotes the
covariant derivative. In a FLRW spacetime, it assumes the form
\begin{equation} \label{KG}
3H\frac{\partial\phi}{\partial t}+\frac{\partial^2\phi}{\partial
t^2}-\nabla^2\phi=0.
\end{equation}

Since the space is isotropic, let us consider a plane wave solution
propagating in the radial direction. As for any plane wave, the
wavelength is supposed very small compared to the cosmological
scale. Therefore, $H$ changes very slowly compared to the wave
function, and the solution has the form
\begin{equation} \label{phi}
\phi \approx \phi_0\, e^{-br} e^{-i(\omega t-kr)},
\end{equation}
with
\begin{equation} \label{k}
k = \frac{\sqrt{2}\omega}{2}\left[1+\sqrt{1+\left(
\frac{3H}{\omega}\right)^2}\right]^{1/2},
\end{equation}
\begin{equation} \label{b}
b =
\frac{3\sqrt{2}H}{2}\left[1+\sqrt{1+\left(\frac{3H}{\omega}\right)^2}\right]^{-1/2}.
\end{equation}
As one can see, the wave amplitude decreases with $r$, with a depth
length equal to $b^{-1}$.

The energy-momentum tensor of this scalar field is
\begin{equation} \label{T}
T^{\nu}_{\mu}=\partial_{\mu}\phi\,
\partial^{\nu}\phi^{\dagger}-\frac{1}{2}\,\partial_{\sigma}\phi\,
\partial^{\sigma}\phi^{\dagger}\, \delta^{\nu}_{\mu}.
\end{equation}
Taking its time component, and using (\ref{phi}), we derive
\begin{equation} \label{rho}
\rho \approx \left(\omega^2+b^2\right)\phi\phi^{\dagger}.
\end{equation}
One can interpret this result by saying that the scalar particle
performs, superposed to the mode of frequency $\omega$, a thermal
motion of temperature $b$.

On this basis, we may evaluate the vacuum fluctuations by doing the
shift $m \rightarrow m+b$ in (\ref{Lambdareg}). The dominant
contributions to the integral will be given by modes with $\omega
\approx m + b$. For $b << m$, the dominance occurs for $\omega
\approx m$, which implies, through (\ref{b}), that $b \approx H$.
For $b \sim m$ or $b>>m$, the dominance occurs for $\omega\approx
b$, leading again, through (\ref{b}), to $b \approx H$. In this
case, however, the plane-wave approximation (\ref{phi})-(\ref{b})
cannot be used anymore, and the identity between $b$ and $H$ will be
taken as an {\it ad hoc} assumption.

Then, after subtracting $\Lambda_0$, we obtain
\begin{equation}\label{Lambda}
\Lambda \approx (m+H)^4-m^4.
\end{equation}
This has a similar structure as in the Casimir effect. Actually, for
$H>>m$ we have the same cutoff-independent result, $\Lambda \approx
H^4$, with $H^{-1}$ playing the role of a distance between Casimir
plates.

Therefore, in the limit of very early times, the cosmological term
scales as $\Lambda \approx H^4$, while for later times ($H<<m$) it
scales as $\Lambda \approx m^3 H$ (we should, however, be careful
with this last conclusion, as discussed below). Let us investigate
the corresponding cosmological scenarios. For simplicity, we will
only consider the spatially flat case.

Leading the Friedmann equation $\rho_T = 3H^2$ into the
conservation equation (\ref{continuidade}), using for matter the
equation of state of radiation, $p_m = \rho_m/3$, and taking for
the vacuum our early-time result $\Lambda = 3H^4$ (the constant
factor is not important, being taken three for convenience), we
obtain the evolution equation
\begin{equation}\label{evolucao}
\dot{H}+2H^2-2H^4=0.
\end{equation}
Apart an integration constant which determines the origin of time,
its solution is
\begin{equation}\label{H}
2t=\frac{1}{H}-\tanh^{-1}H.
\end{equation}

The evolution of $H$ is plotted in Figure 1. As one sees, this
universe has no initial singularity, existing since an infinite
past, when $H$ approaches asymptotically the Planck value $H=1$.
During an infinitely long period we have a quasi-de Sitter,
inflationary expansion, with $H\approx1$. But at a given time
(chosen around $t=0$) we have a huge phase transition, with a
characteristic time scale of a few Planck times, during which $H$
(and so $\Lambda$) falls to nearly zero.

The transition can also be understood in terms of the energy
content. The energy density of radiation is $\rho_m = \rho_T -
\Lambda$, and its relative energy density is $\Omega_m = 1 - H^2$.
Therefore, the transition leads from an empty, vacuum-dominated
universe to a radiation-dominated phase, with $\Omega_m$ approaching
$1$ asymptotically (see Figure 2, where we also plot the relative
energy density of vacuum, $\Omega_{\Lambda}=H^2$). This behavior can
also be described with help of the deceleration factor we obtain
from (\ref{H}). It is $q = 1 - 2H^2$, and suddenly changes from
$-1$, in the quasi-de Sitter phase, to $1$, in the radiation one
(Figure 3).

Let us now consider the limit of late times, for which
$\Lambda=\sigma H$, with $\sigma\approx m^3$. We have shown
elsewhere \cite{Borges} that, in the radiation phase, $a\propto
t^{1/2}$, with $\rho_m = 3/(4t^2)$, as in the standard model. On the
other hand, in the dust phase we have
\begin{equation}\label{a}
a = C\left( e^{\sigma t/2}-1\right)^{2/3},
\end{equation}
where $C$ is an integration constant.

For early times ($\sigma t<<1$), we have $a\propto t^{2/3}$, as in
the Einstein-de Sitter model. This decelerated phase follows until
very recently, when vacuum begins dominating again and the expansion
reenters in an accelerated phase, which, as one can see from
(\ref{a}), tends asymptotically to a de Sitter universe. We have
analyzed the redshift-distance relation for supernovas Ia in this
model, obtaining a fit of observational data as good as in the
$\Lambda$CDM model. The obtained present values of $H$ and
$\Omega_m$ and the universe age are also in good accordance with
other observations \cite{Borges}. Finally, the analysis of evolution
of density perturbations until the present time shows no important
difference compared to the $\Lambda$CDM model.

In the de Sitter limit, the Hubble parameter is given, as we know,
by $H=\sqrt{\Lambda/3}$. Therefore, using $\Lambda \approx m^3 H$,
one derives the results $H\approx m^3$ and $\Lambda\approx m^6$. The
former is an expression of the famous Eddington-Dirac large number
coincidence, provided we take $m$ of the order of the pion mass,
i.e., the order of the energy scale of the QCD chiral transition, as
initially supposed. The last relation, on the other hand, was
suggested by Zel'dovich four decades ago, on the basis of different
arguments.

Nevertheless, we should be careful before concluding that the
present universe evolves as described above. Our approach is based
on a macroscopic, semi-classical reasoning, and we still do not have
a microscopic description of vacuum decay. At late times the decay
probably depends on the mass of the produced particles, and so we
have no guarantee that vacuum is still decaying. If it stopped
decaying at some earlier time, we just have, after the primordial
transition, a $\Lambda$CDM universe.

To conclude, some words about the entropy of this universe. If the
vacuum fluctuations are thermally distributed, as suggested by
(\ref{Lambdareg}), the number of states inside a volume $V$ may be
estimated as
\begin{equation}\label{N}
N \approx V \int_0^{\infty}
\frac{\omega^2\,d\omega}{e^{\omega/(m+H)}-1}.
\end{equation}

For late times we have $H<<m$, and $N\approx Vm^3$. In the de Sitter
limit, taking $V$ as the Hubble volume, and $m^3\approx H$, we
obtain $N\approx H^{-2}\approx 10^{120}$. That is, in the final de
Sitter phase, the entropy inside the Hubble sphere is equal to the
area of its surface, which is an expression of the holographic
conjecture \cite{Bousso,GRF2003}.

On the other hand, during the primeval quasi-de Sitter phase, we
have $H>>m$, and (\ref{N}) leads to $N\approx VH^3$. Now, by taking
the Hubble volume we obtain $N\approx 1$, which is, again, equal to
the area of the Hubble surface. In this way, one may conclude that
the primordial phase transition leads a universe of very low entropy
into a state of very high entropy. The thermodynamic and time arrows
coincide.

${ }$

I am indebt to R. Abramo, A. Saa, I. Shapiro and C. Pigozzo for
useful discussions and assistance.

\vspace{1.5cm}

\begin{figure}[hb]
\begin{center}
\includegraphics[height=6cm,width=10cm]{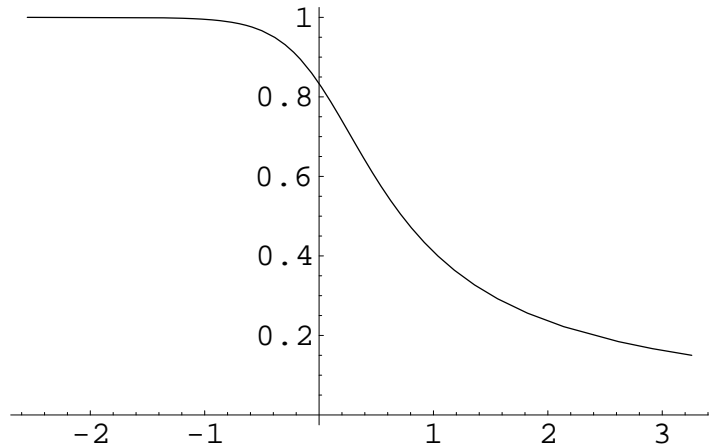}
\end{center}
\caption{The Hubble parameter as a function of time (in Planck
units)}
\end{figure}

\begin{figure}[h]
\begin{center}
\includegraphics[height=6cm,width=10cm]{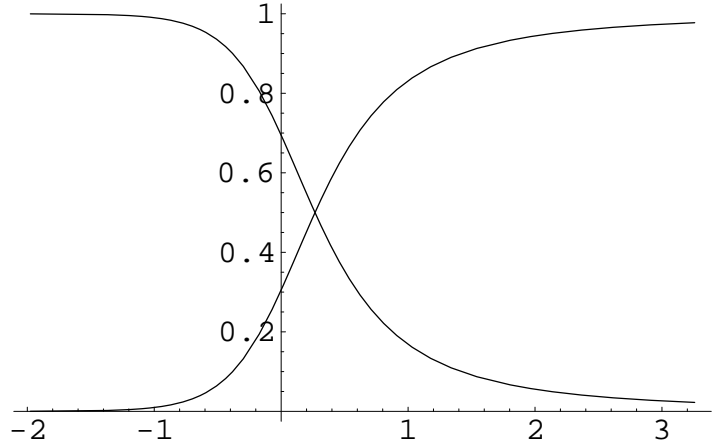}
\end{center}
\caption{The relative energy densities of radiation and vacuum as
functions of time}
\end{figure}

\begin{figure}[hb]
\begin{center}
\includegraphics[height=6cm,width=10cm]{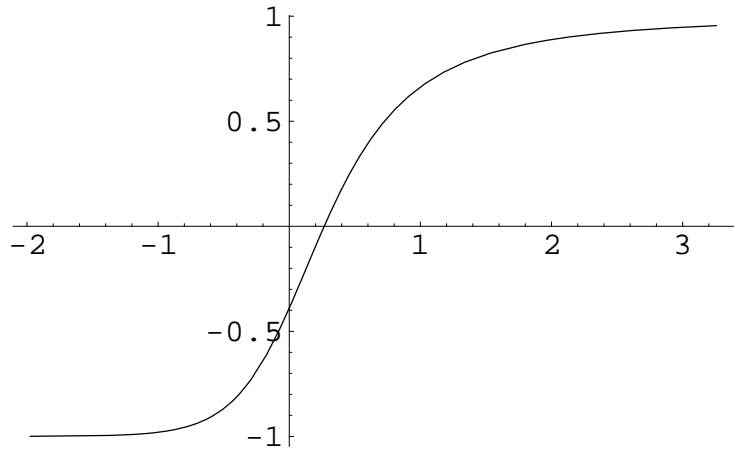}
\end{center}
\caption{The deceleration parameter as a function of time}
\end{figure}

\end{document}